\documentclass{aa}
\usepackage[dvips]{graphicx}

\begin{document}

\title{Relativistic corrections to the Sunyaev-Zeldovich effect for clusters of galaxies: Effect of the motion of the observer}

\author{Satoshi Nozawa\inst{1} \and Naoki Itoh\inst{2} \and Yasuharu Kohyama\inst{3}}

\offprints{Naoki Itoh}

\institute{Josai Junior College for Women, 1-1 Keyakidai, Sakado-shi, Saitama,
           350-0290, Japan \\
           e-mail: snozawa@josai.ac.jp \\
     \and
           Department of Physics, Sophia University, 7-1 Kioi-cho, Chiyoda-ku,
           Tokyo, 102-8554, Japan \\
           email: n\_itoh@sophia.ac.jp \\
     \and
           Mizuho Information and Research Institute, 2-3 Kanda-Nishiki-cho,
           Chiyoda-ku, Tokyo, 101-8443, Japan \\
           yasuharu.kohyama@gene.mizuho-ir.co.jp
           }

\date{Received  \hspace{4.0cm} / Accepted \hspace{4.0cm} }

\authorrunning{Nozawa et al.}
\titlerunning{Relativistic Corrections to the Sunyaev-Zeldovich Effect}


\abstract{
We extend the formalism of the relativistic thermal and kinematical Sunyaev-Zeldovich effects to the observer's system (the Solar System) moving with a velocity $\vec{\beta}_{S} \equiv \vec{v}_{S}/c$ with respect to the cosmic microwave background radiation.  The present formulation makes full use of the Lorentz covariance properties of the problem and gives solutions with high precision.  We confirm the results recently obtained by Chluba, Huetsi, and Sunyaev in the lowest order of the observer's velocity $\beta_{S}$.  We give a more general analytic expression for the thermal and kinematical Sunyaev-Zeldovich effects corresponding to the observer's system (the Solar System) with the power series expansion approximation in terms of $\theta_{e} \equiv k_{B}T_{e}/mc^{2}$, where $T_{e}$ and $m$ are the electron temperature and the electron mass, respectively.

}

\maketitle

\section{Introduction}

  Compton scattering of the cosmic microwave background radiation (CMBR) by hot intracluster gas --- the Sunyaev-Zeldovich effect (Zeldovich \& Sunyaev 1969; Sunyaev \& Zeldovich 1972, 1980a,b, 1981) --- provides a useful method for studies of cosmology (see recent excellent reviews: Birkinshaw 1999; Carlstrom, Holder, \& Reese 2002).  The original Sunyaev-Zeldovich formula has been derived from a kinetic equation for the photon distribution function taking the Compton scattering by electrons into account: the Kompaneets equation (Kompaneets 1957; Weymann 1965).  The original Kompaneets equation has been derived with a nonrelativistic approximation for the electron; however, recent X-ray observations have revealed the existence of many high-temperature galaxy clusters (Tucker et al. 1998; Markevitch 1998; Allen et al. 2001; Schmidt et al. 2001; Allen et al. 2002).  Some galaxy clusters have been found to possess the electron temperature $k_{B} T_{e} \simeq $20keV.  Wright (1979) and Rephaeli and his collaborator (Rephaeli 1995; Rephaeli \& Yankovitch 1997) have done pioneering work taking the relativistic corrections to the Sunyaev-Zeldovich effect into account for clusters of galaxies.

  In recent years remarkable progress has been made in theoretical studies of the relativistic corrections to the Sunyaev-Zeldovich effects for clusters of galaxies.  Stebbins (1997) generalized the Kompaneets equation.  Itoh, Kohyama \& Nozawa (1998) have adopted a relativistically covariant formalism to describe the Compton scattering process (Berestetskii, Lifshitz, \& Pitaevskii 1982; Buchler \& Yueh 1976), thereby obtaining higher-order relativistic corrections to the thermal Sunyaev-Zeldovich effect in the form of the Fokker-Planck expansion.  In their derivation, the scheme to conserve the photon number at every stage of the expansion proposed by Challinor \& Lasenby (1998) played an essential role.

  Nozawa, Itoh \& Kohyama (1998b) have extended their method to the case where the galaxy cluster is moving with a peculiar velocity with respect to CMBR.  They have thereby obtained the relativistic corrections to the kinematical Sunyaev-Zeldovich effect.  Challinor \& Lasenby (1999b) then confirmed the correctness of the result obtained by Nozawa et al. (1998b).  Sazonov \& Sunyaev (1998a,b) calculated the kinematical Sunyaev-Zeldovich effect by a different method.

  Itoh, Nozawa \& Kohyama (2000) have also applied their method to calculation of the relativistic corrections to the polarization Sunyaev-Zeldovich effect (Sunyaev \& Zeldovich 1980b, 1981).  They thereby confirmed the result of Challinor, Ford \& Lasenby (1999a) which was obtained with a completely different method.  Recent work on the polarization Sunyaev-Zeldovich effect include Audit \& Simons (1999), Hansen \& Lilje (1999), and Sazonov \& Sunyaev (1999).

  Sazonov \& Sunyaev (1998a,b) reported the results of the Monte Carlo calculations on the relativistic corrections to the Sunyaev-Zeldovich effect.  In Sazonov \& Sunyaev (1998b), a numerical table that summarizes the results of the Monte Carlo calculations was presented.  Nozawa et al. (2000) presented an accurate analyic fitting formula that has a high accuracy for the ranges $0.00 \leq \theta_{e} \leq 0.05$ and $0 \leq X \leq 20$, where $\theta_{e} \equiv k_{B}T_{e}/m_{e}c^{2}$, $X \equiv \hbar \omega/k_{B}T_{0}$.  Another fitting formula with a still higher precision that is valid for the more limited ranges $0.00 \leq \theta_{e} \leq 0.035$, $0 \leq X \leq 15$ was developed by Itoh et al. (2000b).  Relativistic corrections to the double scattering effect on the Sunyaev-Zeldovich effect was calculated by Itoh et al. (2001).  Dolgov et al. (2001) carried out an independent numerical calculation, and their result shows excellent agreement with that of Itoh et al. (2001) for the case of small optical depth.

  Very recently Kitayama et al. (2004) measured the Sunyaev-Zeldovich effect in the galaxy cluster RX J1347-1145.  They have discovered that the south-east excess component of this galaxy cluster has a temperature $k_{B}T_{ex}=28.5 \pm 7.3$keV.  They attribute the high temperature of this excess component to a recent major merger as discussed by Sarazin (2003).  Therefore it is important to present accurate numerical data for $k_{B}T_{e} \geq 25$keV.  Itoh \& Nozawa (2004) have accordingly presented an accurate numerical table for the relativistic corrections to the Sunyaev-Zeldovich effect for clusters of galaxies in the ranges $0.002 \leq \theta_{e} \leq 0.100$.  For analyses of the galaxy clusters with extremely high temperatures, the results of the calculation of the relativistic thermal bremsstrahlung Gaunt factor (Nozawa, Itoh \& Kohyama 1998a) and their accurate analytic fitting formulae (Itoh et al. 2000b) will be useful.  The nonrelativistic electron-electron thermal bremsstrahlung Gaunt factor has been also calculated by Itoh, Kawana \& Nozawa (2002a).

  It now appears that all the necessary theoretical tools are in place for accurate analysis of the observational data of the Sunyaev-Zeldovich effect for galaxy clusters.  Recently, however, Chluba, Huetsi, \& Sunyaev (2005) point out the importance of the influence of the Solar System's motion with respect to the CMBR rest frame.  Assuming that the CMBR dipole is fully motion-induced, we deduce that the Solar System is moving with a velocity $\beta_{S} \equiv v_{S}/c$ = 1.241$\times10^{-3}$ towards the direction $(\ell, b)=(264.14^{\circ} \pm 0.15^{\circ}$, $48.26^{\circ} \pm 0.15^{\circ}$) (Smoot et al. 1977; Fixsen et al. 1996; Fixsen \& Mather 2002).  Chluba et al. (2005) have calculated corrections to the Sunyaev-Zeldovich effect arising from the motion of the Solar System.  In this paper we wish to calculate those corrections to the Sunyaev-Zeldovich effect which are due to the motion of the Solar System in a more general way.  As a matter of fact, it turns out that one can make full use of the Lorentz covariance properties of the problem and obtain solutions with high precision by extending the results obtained by Itoh, Kohyama \& Nozawa (1998) and Nozawa, Itoh \& Kohyama (1998b).

  The present paper is organized as follows.  In $\S$ 2, the general formalism is presented in deriving the thermal and kinematical Sunyaev-Zeldovich effects by taking the motion of the Solar System into account.  With the power series expansion approximation, an analytic expression including the relativistic effects is derived for the thermal and kinematical Sunyaev-Zeldovich effects also by taking the motion of the Solar System into account.  Numerical results are presented in $\S$ 3, and finally, concluding remarks are given in $\S$ 4.

\section{Lorentz-boosted Kompaneets equation}

  In the present section we extend the work of Nozawa, Itoh, \& Kohyama (1998b) by first extending the Kompaneets equation to a system (the cluster of galaxies) moving with a peculiar velocity with respect to the CMBR.  To formulate the kinetic equation for the photon distribution function we use a relativistically covariant formalism (Berestetskii, Lifshitz, \& Pitaevskii 1982; Buchler \& Yueh 1976).  As a reference system, we have chosen the system which is fixed to the cosmic  microwave background radiation (CMBR).  The $z$-axis is fixed to a line connecting the observer and the center of mass of the cluster of galaxies (CG), first assuming that the observer is fixed to the CMBR frame.  We fix the positive direction of the $z$-axis as the direction of the propagation of a photon from the cluster to the observer.  In this reference system, the center of mass of the CG is moving with a peculiar velocity $\vec{\beta} (\equiv \vec{v}/c$) with respect to the CMBR.  For simplicity, we choose the direction of the velocity in the $x$-$z$ plane, i.e. $\vec{\beta} = (\beta_{x}, 0, \beta_{z})$.

  In the CMBR frame the four-momenta of the initial electron and photon are $p = (E, \vec{p})$ and $k = (\omega, 0, 0, k)$, respectively, while the four-momenta of the final electron and photon are $p^{\prime} = (E^{\prime}, \vec{p}^{\prime})$ and $k^{\prime} = (\omega^{\prime}, \vec{k}^{\prime})$, respectively.  The electron distribution functions in the initial and final states are Fermi--like in the CG frame.  They are related to the electron distribution functions in the CMBR frame as follows (Landau \& Lifshitz 1975):
\begin{eqnarray}
f(E) & = &  f_{C}(E_{C})  \, , \\
f(E^{\prime}) & = &  f_{C}(E_{C}^{\prime})  \,  ,  \\
E_{C} & = & \gamma \, \left(E - \vec{\beta} \cdot  \vec{p} \right) \, ,   \\
E_{C}^{\prime} & = & \gamma \, \left(E^{\prime} - \vec{\beta} \cdot \vec{p}^{\prime} \right) \, ,  \\
\gamma & \equiv & \frac{1}{\sqrt{1 - \beta^{2}}}   \, ,
\end{eqnarray}
where the suffix $C$ denotes the CG frame.

In addition to $\theta_{e}$, there is another parameter $\vec{\beta}$.  For most of the CG, $\beta \ll 1$ is realized.  For example, $\beta \approx $ 1/300 for a typical value of the peculiar velocity $v$=1,000km/s.  Therefore it should be sufficient to expand in powers of $\beta$ and to retain up to  $O(\beta^{2})$ contributions.  We assume the initial photon distribution of the CMBR to be Planckian with a temperature $T_{0}$:
\begin{equation}
n_{0} (X) \, = \, \frac{1}{e^{X} - 1} \, , 
\end{equation}
where
\begin{equation}
X \, \equiv \, \frac{\omega}{k_{B} T_{0}}  \, .
\end{equation}

  Nozawa, Itoh \& Kohyama (1998b) have obtained the following expression for the fractional distortion of the photon spectrum:

\begin{eqnarray}
\frac{\Delta n(X)}{n_{0}(X)} & = & \frac{\tau \, X e^{X}}{e^{X}-1} \, \theta_{e} \, \left[  \, \,
Y_{0} \, + \, \theta_{e} Y_{1} \, + \, \theta_{e}^{2} Y_{2} \right. \, \nonumber \\
&  & \left. \, + \,  \theta_{e}^{3} Y_{3} \, + \, \theta_{e}^{4} Y_{4} \,  \right]  \,   \nonumber  \\
  & + & \frac{\tau \, X e^{X}}{e^{X}-1}  \, \beta^{2} \, \left[ \, \, \frac{1}{3} Y_{0} \, + \, \theta_{e} \left( \, \frac{5}{6} Y_{0} \, + \, \frac{2}{3} Y_{1} \, \right) \, \right]    \, \nonumber \\
  & + & \frac{\tau \, X e^{X}}{e^{X}-1}  \, \beta \, P_{1}(\hat{\beta}_{z}) \, \left[ \, \, 1 \, + \, \theta_{e} C_{1} \, + \, \theta_{e}^{2} C_{2} \,  \right]    \, \nonumber \\
& + & \frac{\tau \, X e^{X}}{e^{X}-1} \beta^{2}  P_{2} (\hat{\beta}_{z}) \, \left[ \, D_{0} \, + \, \theta_{e} D_{1} \, \right] \, ,  \\
\hat{\beta}_{z} & \equiv & \frac{\beta_{z}}{\beta} \, = \, {\rm cos} \theta_{\gamma}  \, , \\
P_{1} (\hat{\beta}_{z}) & = & \hat{\beta}_{z} \, , \\
P_{2} (\hat{\beta}_{z}) & = & \frac{1}{2} \, \left( 3 \hat{\beta}_{z}^{2} - 1 \right) \, ,
\end{eqnarray}
where $\theta_{\gamma}$ is the angle between the directions of the peculiar velocity of the cluster ($\vec{\beta}$) and the initial photon momentum ($\vec{k}$), which is chosen as the positive $z$-direction.  The reader should notice that this sign convention for the positive $z$-direction is the opposite of the ordinary one.  Thus a cluster moving away from the observer has $\hat{\beta}_{z} < 0$.  However, because of the positive sign in front of the $P_{1}(\hat{\beta}_{z})$ term, one obtains $\Delta n(X) < 0$ in this case, as one should.  The coefficients are defined as follows:
\begin{eqnarray}
Y_{0} & = & - 4 \, + \tilde{X}  \,  , \\
Y_{1} & = & - 10 + \frac{47}{2} \tilde{X} - \frac{42}{5} \tilde{X}^{2} + \frac{7}{10} \tilde{X}^{3}  \nonumber \\
& + &  \tilde{S}^{2} \left( - \frac{21}{5} + \frac{7}{5} \tilde{X} \right) \,  ,  \\
Y_{2} & = & - \frac{15}{2} + \frac{1023}{8} \tilde{X} - \frac{868}{5} \tilde{X}^{2} + \frac{329}{5} \tilde{X}^{3} \nonumber \\
& & - \frac{44}{5} \tilde{X}^{4} + \frac{11}{30} \tilde{X}^{5}  \nonumber \\ 
& + &  \tilde{S}^{2} \left( - \frac{434}{5} + \frac{658}{5} \tilde{X}  - \frac{242}{5}  \tilde{X}^{2} + \frac{143}{30} \tilde{X}^{3} \right) \nonumber \\
& + &  \tilde{S}^{4} \left( - \frac{44}{5} + \frac{187}{60} \tilde{X} \right) \, , 
\end{eqnarray}
\begin{eqnarray}
Y_{3} & = & \frac{15}{2} + \frac{2505}{8} \tilde{X} - \frac{7098}{5} \tilde{X}^{2} + \frac{14253}{10} \tilde{X}^{3}  \nonumber  \\
&  & - \frac{18594}{35} \tilde{X}^{4} + \frac{12059}{140} \tilde{X}^{5} - \frac{128}{21} \tilde{X}^{6} + \frac{16}{105} \tilde{X}^{7} \nonumber \\ 
& + &  \tilde{S}^{2} \left( - \frac{7098}{10} + \frac{14253}{5} \tilde{X} - \frac{102267}{35}  \tilde{X}^{2}  \right. \nonumber \\
& & \left. + \frac{156767}{140} \tilde{X}^{3} - \frac{1216}{7}  \tilde{X}^{4} + \frac{64}{7} \tilde{X}^{5} \right)  \nonumber  \\
& + &  \tilde{S}^{4} \left( - \frac{18594}{35} + \frac{205003}{280} \tilde{X} - \frac{1920}{7}  \tilde{X}^{2} \right. \nonumber  \\
& & \left.  + \frac{1024}{35} \tilde{X}^{3} \right) \nonumber  \\
& + &  \tilde{S}^{6} \left( - \frac{544}{21} + \frac{992}{105} \tilde{X} \right) \, , \\
Y_{4} & = & - \frac{135}{32} + \frac{30375}{128} \tilde{X} - \frac{62391}{10} \tilde{X}^{2} + \frac{614727}{40} \tilde{X}^{3}  \nonumber  \\
&  &  - \frac{124389}{10} \tilde{X}^{4} + \frac{355703}{80} \tilde{X}^{5} - \frac{16568}{21} \tilde{X}^{6}  \nonumber \\ 
& &  + \frac{7516}{105} \tilde{X}^{7} - \frac{22}{7} \tilde{X}^{8} + \frac{11}{210} \tilde{X}^{9} \nonumber \\
& + &  \tilde{S}^{2} \left( - \frac{62391}{20} + \frac{614727}{20} \tilde{X} - \frac{1368279}{20} \tilde{X}^{2} \right. \nonumber  \\
&  &   + \frac{4624139}{80} \tilde{X}^{3} - \frac{157396}{7} \tilde{X}^{4} + \frac{30064}{7} \tilde{X}^{5} \nonumber \\
& & \left. - \frac{2717}{7} \tilde{X}^{6} + \frac{2761}{210} \tilde{X}^{7}   \right)  \nonumber  \\
& + &  \tilde{S}^{4} \left( - \frac{124389}{10} + \frac{6046951}{160} \tilde{X} - \frac{248520}{7} \tilde{X}^{2}   \right. \nonumber  \\
&  &  \left. + \frac{481024}{35} \tilde{X}^{3} - \frac{15972}{7} \tilde{X}^{4} + \frac{18689}{140} \tilde{X}^{5}  \right) \nonumber  \\
& + &  \tilde{S}^{6} \left( - \frac{70414}{21} + \frac{465992}{105} \tilde{X} - \frac{11792}{7} \tilde{X}^{2} \right. \nonumber  \\
& & \left. + \frac{19778}{105} \tilde{X}^{3} \right) \nonumber \\
& + &  \tilde{S}^{8} \left( - \frac{682}{7} + \frac{7601}{210} \tilde{X} \right) \, ,
\end{eqnarray}
\begin{eqnarray}
C_{1} & = & 10 - \frac{47}{5} \tilde{X} + \frac{7}{5} \tilde{X}^{2} + \frac{7}{10} \tilde{S}^{2}  \,  ,  \\
C_{2} & = & 25 - \frac{1117}{10} \tilde{X} + \frac{847}{10} \tilde{X}^{2} - \frac{183}{10} \tilde{X}^{3} + \frac{11}{10} \tilde{X}^{4}    \nonumber \\ 
& & + \tilde{S}^{2} \left( \frac{847}{20} - \frac{183}{5} \tilde{X}  + \frac{121}{20}  \tilde{X}^{2} \right)  +  \frac{11}{10} \tilde{S}^{4}  \, ,   \\
\nonumber \\
D_{0} & = & - \frac{2}{3} + \frac{11}{30} \tilde{X} \, , \\
D_{1} & = & - 4 + 12 \tilde{X} - 6 \tilde{X}^{2} + \frac{19}{30} \tilde{X}^{3}    \nonumber \\
& + & \tilde{S}^{2} \left( - 3 + \frac{19}{15} \tilde{X} \right) \,  ,
\end{eqnarray}
and
\begin{eqnarray}
\tau & \equiv & \sigma_{T} \int d \ell N_{e}  \, , \\
\tilde{X} & \equiv &  X \, {\rm coth} \left( \frac{X}{2} \right)  \, , \\
\tilde{S} & \equiv & \frac{X}{ \displaystyle{ {\rm sinh} \left( \frac{X}{2} \right)} }   \, ,
\end{eqnarray}
where $N_{e}$ is the electron number density in the CG frame, the integral in Eq. (21) is over the photon path length in the cluster, and $\sigma_{T}$ is the Thomson cross section.

  In the next step of the calculation, we suppose that the observer's system (the Solar System) is moving with a velocity $\vec{\beta}_{S}(\equiv \vec{v}_{S}/c)$ with respect to the CMBR.  The photon distribution function in the CMBR system $n(\omega, \vec{k})$ and the photon distribution function in the Solar System $n_{S}(\omega_{S}, \vec{k}_{S})$ are related by (Landau \& Lifshitz 1975)
\begin{eqnarray}
n(\omega, \vec{k}) & = & n_{S}(\omega_{S}, \vec{k}_{S}) \, , \\
\omega & = & \gamma_{S} ( \omega_{S} + \vec{\beta}_{S} \cdot \vec{k}_{S} ) \, ,  \\
\gamma_{S} & \equiv & \frac{1}{ \sqrt{1 - \beta_{S}^{2} }} \, .
\end{eqnarray}
The photon unit wave vector in the Solar System $\hat{k}_{S}$ and the photon unit wave vector in the CMBR system $\hat{k}$ are related by (M{\rm $\o$}ller 1962)\begin{eqnarray}
\hat{k} & = & \left(\frac{ \hat{k}_{S} \cdot \hat{\beta}_{S} + \beta_{S} }{1 + \hat{k}_{S} \cdot \vec{\beta}_{S} } \right) \hat{\beta}_{S} +   \frac{ \hat{k}_{S} - ( \hat{k}_{S} \cdot \hat{\beta}_{S} ) \hat{\beta}_{S} }{ \gamma_{S}(1 + \hat{k}_{S} \cdot \vec{\beta}_{S} ) }   \, ,
\end{eqnarray}
where $\hat{\beta}_{S}$ is a unit vector in the direction of $\vec{\beta}_{S}$.

  Therefore, our task is to rewrite Eq. (8) for the expression corresponding to the Solar System.  The isotropic Planck distribution in the CMBR system, Eq. (6), corresponds to the distribution including the dipolar distortion in the Solar System
\begin{eqnarray}
n_{S}^{0}(X_{S}, \hat{k}_{S}) & = & \frac{1}{ {\rm exp} \left\{ \gamma_{S} X_{S} (1 + \beta_{S} \mu_{S}) \right\} - 1 }  \, ,
\end{eqnarray}
where
\begin{eqnarray}
X_{S} & \equiv & \frac{\omega_{S}}{k_{B}T_{0}}  \, ,  \\
\mu_{S} & \equiv & {\rm cos} \theta_{S}  \, ,
\end{eqnarray}
$\theta_{S}$ being the angle between $\vec{\beta}_{S}$ and $\vec{k}_{S}$.
Here we have used the relationship
\begin{eqnarray}
X & = & \gamma_{S} X_{S}(1 + \beta_{S} \mu_{S}) \, .
\end{eqnarray}
The velocity vecor $\vec{\beta}$ of the galaxy cluster in the CMBR system is transformed to the velocity vector $\vec{\beta}^{\prime}$ of the galaxy cluster in the Solar System by the relationship (M{\rm $\o$}ller 1962)
\begin{eqnarray}
\vec{\beta} & = & \left(\frac{ \vec{\beta}^{\prime} \cdot \hat{\beta}_{S} + \beta_{S} }{1 + \vec{\beta}^{\prime} \cdot \vec{\beta}_{S} } \right) \hat{\beta}_{S} +   \frac{ \vec{\beta}^{\prime} - ( \vec{\beta}^{\prime} \cdot \hat{\beta}_{S}  ) \hat{\beta}_{S} }{ \gamma_{S}(1 + \vec{\beta}^{\prime} \cdot \vec{\beta}_{S} ) }   \, .
\end{eqnarray}
Therefore, $\beta_{z}$ in the CMBR system in Eq. (9) is transformed to
\begin{eqnarray}
\beta_{z} & \equiv & \vec{\beta} \cdot \hat{k}  \nonumber \\
& = & \frac{1}{(1 + \vec{\beta}^{\prime} \cdot \vec{\beta}_{S})(1 + \hat{k}_{S} \cdot \vec{\beta}_{S})} \, \times \nonumber \\
& & \left[ (\vec{\beta}^{\prime} \cdot \hat{\beta}_{S} + \beta_{S})(\hat{k}_{S} \cdot \hat{\beta}_{S} + \beta_{S}) \right. \nonumber \\
& & \left. + (1 - \beta_{S}^{2}) \left\{ \vec{\beta}^{\prime} \cdot \hat{k}_{S} - (\vec{\beta}^{\prime} \cdot \hat{\beta}_{S})(\hat{\beta}_{S} \cdot \hat{k}_{S}) \right\} \right]  \, .
\end{eqnarray}

  We write the photon distribution function in the Solar System frame that has a distortion caused by the thermal and kinematical Sunyaev-Zeldovich effects as
\begin{eqnarray}
n_{S}(X_{S},\hat{k}_{S}) & = & n_{S}^{0}(X_{S},\hat{k}_{S}) + \Delta n_{S}(X_{S},\hat{k}_{S}) \, .
\end{eqnarray}
Here $\Delta n_{S}(X_{S},\hat{k}_{S})$ is given by
\begin{eqnarray}
\Delta n_{S}(X_{S},\hat{k}_{S}) & = & \frac{\tau \, X e^{X}}{(e^{X}-1)^{2}} \left\{ \, \theta_{e} \, \left[  \, \,
Y_{0} \, + \, \theta_{e} Y_{1} \, + \, \theta_{e}^{2} Y_{2} \right. \right. \nonumber \\
&  & \left. \, + \, \theta_{e}^{3} Y_{3} \, + \, \theta_{e}^{4} Y_{4} \,  \right] \,   \nonumber  \\
  & + &  \beta^{2} \, \left[ \, \, \frac{1}{3} Y_{0} \, + \, \theta_{e} \left( \, \frac{5}{6} Y_{0} \, + \, \frac{2}{3} Y_{1} \, \right) \, \right]    \, \nonumber \\ 
  & + &  \beta \, P_{1}(\hat{\beta}_{z}) \, \left[ \, \, 1 \, + \, \theta_{e} C_{1} \, + \, \theta_{e}^{2} C_{2} \,  \right]    \, \nonumber \\
& + & \left.  \beta^{2}  P_{2} (\hat{\beta}_{z}) \, \left[ \, D_{0} \, + \, \theta_{e} D_{1} \, \right] \right\} \, , 
\end{eqnarray}
where one should rewrite $X$ by using Eq. (31) and also $\beta P_{1}(\hat{\beta}_{z})$ and $\beta^{2} P_{2}(\hat{\beta}_{z})$ by using Eqs. (32), (33).  The present result can be used in order to analyze the observational data of the Sunyaev-Zeldovich effect.  We note that $\beta_{S} = 1.241 \times 10^{-3}$ is a known quantity and $\mu_{S}$= cos$\theta_{S}$ is also a known quantity for the galaxy cluster under consideration.  For present and near-future observations of the Sunyaev-Zeldovich effect, $\beta^{2}$ terms in Eq. (35) are beyond observational accuracy, such that we only consider the first and third terms in the braces of Eq. (35).  When the optical depth of the galaxy cluster $\tau$ is known, we obtain the best-fit values of $\theta_{e}$ and $\beta_{z}$ by fitting the observational data with the first and third terms of Eq. (35) written for the Solar System frame.  In this way we can obtain the cluster temperature, as well as the cluster velocity in the direction of the photon propagation vector observed by the observer fixed to the CMBR system.

  In the somewhat more distant future accuracy in the observation of the Sunyaev-Zeldovich effect will be hopefully improved dramatically.  In that case one should also include the $\beta^{2}$ terms in Eq. (35) and use the full expression for analysis of the observational data.  In doing so it should be possible to also determine the cluster velocity perpendicular to the photon propagation vector.
  We define the distortion of the spectral intensity in the Solar System frame as follows:
\begin{eqnarray}
\Delta I_{S} & \equiv & \Delta n_{S}(X_{S},\hat{k}_{S}) X_{S}^{3} \, .
\end{eqnarray}

\section{Numerical results}

  Equations (31) and (35) present our main results for the effect of the observer's the motion (the Solar System).  It should be emphasized here that because no expansions have been made so far in deriving Eqs. (31) and (35) for $\beta_{S}$, they are general expressions for arbitrary values of $\beta_{S}$.  In practical cases, however, $\beta_{S} \ll 1$ is achieved.  By expanding $X$ around $X_{S}$ and keeping only lowest order terms for both $\beta_{S}$ and $\theta_{e}$, we have reproduced the Eqs. (12a) and (12b) of Chluba, Huetsi, and Sunyaev (2005).

\begin{figure}
\begin{center}
\includegraphics[angle=-90,scale=0.35]{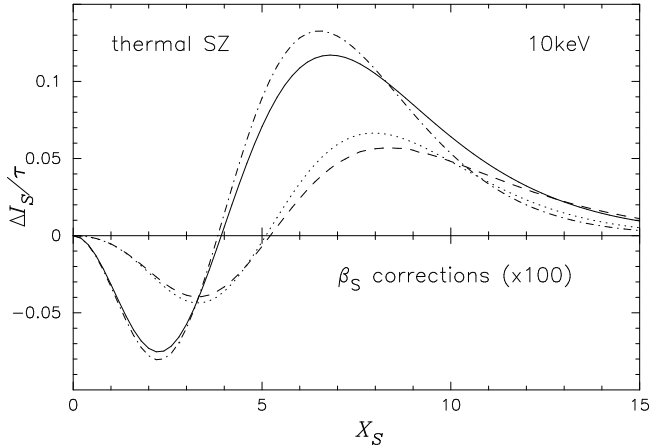}
\end{center}
\caption{Spectral intensity of the thermal Sunyavev-Zeldovich effect for $k_{B}T_{e}=10$keV and $\mu_{S}=-1$ as a function of $X_{S}$.  The solid curve is the full contribution.  The dash-dotted curve is the lowest-order ($Y_{0}$) contribution.  The dashed curve is the extraction of the full $\beta_{S}$ corrections (multiplied by 100).  The dotted curve shows results of Chluba et al. (2005) (multiplied by 100).  Refer to the main text for detailed explanations.}
\end{figure}

  In Fig. 1, we plot the spectral intensity of the thermal Sunyavev-Zeldovich effect for $k_{B}T_{e}=10$keV and $\mu_{S}=-1$ as a function of $X_{S}$.  Here the solid curve is the full contribution ($Y_{0}$ to $Y_{4}$ terms) of the first line of Eq. (35), and the dash-dotted curve is the lowest-order ($Y_{0}$ term) contribution of the first line of Eq. (35).  The dashed curve is the extraction of the full $\beta_{S}$ corrections of the first line of Eq. (35), which is multiplied by 100 in order to be visible in the same figure.  This curve contains full $\beta_{S}$ corrections.  The dotted curve shows the results of Chluba, Huetsi, and Sunyaev (2005) (Eq. (12a) in their paper), where only lowest-order corrections are included for both $\beta_{S}$ and $\theta_{e}$.  This correction is also multiplied by 100 in order to be visible in the same figure.  It should be noted that the $\beta_{S}^{2}$ corrections are totally negligible.  The essential difference in two curves are due to higher-order relativistic corrections of $\theta_{e}$, which is present in Eq. (35) but is absent in Eq. (12a) of Chluba, Huetsi, and Sunyaev (2005).  Therefore, higher order relativistic corrections of $\theta_{e}$ are important in discussing the effect of the observer's motion in the thermal Sunyaev-Zeldovich effect with high precision.

\begin{figure}
\begin{center}
\includegraphics[angle=-90,scale=0.35]{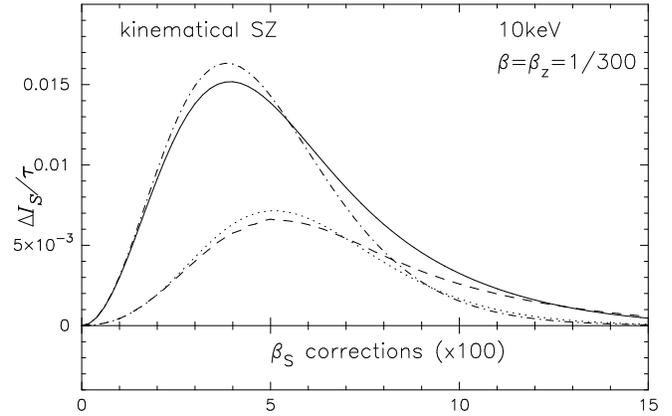}
\end{center}
\caption{Spectral intensity of the kinematical Sunyavev-Zeldovich effect for $k_{B}T_{e}=10$keV, $\beta=\beta_{z}=1/300$, and $\mu_{S}=-1$ as a function of $X_{S}$. The solid curve is the full contribution.  The dash-dotted curve is the lowest-order contribution $\beta P_{1}(\hat{\beta}_{z})$.  The dashed curve is the extraction of the full $\beta_{S}$ corrections (multiplied by 100).  The dotted curve shows the results of Chluba et al. (2005) (multiplied by 100).  Refer to the main text for a detailed explanations.}
\end{figure}

In Fig. 2, we have also plotted the spectral intensity of the kinematical Sunyavev-Zeldovich effect for $k_{B}T_{e}=10$keV, $\beta=\beta_{z}=1/300$, and $\mu_{S}=-1$ as a function of $X_{S}$.  Note that the velocity of the cluster relative to the CMBR and the velocity of the observer relative to the CMBR are parallel in this case.  The solid curve is the full contribution (1 to $C_{2}$ terms) of the third line of Eq. (35), and the dash-dotted curve is the lowest-order contribution of the third line of Eq. (35).  The dashed curve is the extraction of the full $\beta_{S}$ corrections of the third line of Eq. (35), which is multiplied by 100 in order to be visible in the same figure.  This curve contains full $\beta_{S}$ corrections.  The dotted curve shows the results of Chluba, Huetsi, and Sunyaev (2005) (Eq. (12b) in their paper), where only lowest-order corrections are included for both $\beta_{S}$ and $\theta_{e}$.  This correction is also multiplied by 100 in order to be visible in the same figure.  It should also be noted that the $\beta_{S}^{2}$ corrections are totally negligible.  The essential difference in two curves are due to higher-order relativistic corrections of $\theta_{e}$, which is present in Eq. (35) but is absent in Eq. (12b) of Chluba, Huetsi, and Sunyaev (2005).  Therefore, higher order relativistic corrections of $\theta_{e}$ are again important in discussing the effect of the motion of the observer in the kinematical Sunaev-Zeldovich effect with high precision.

\section{Concluding remarks}

 We have calculated the effect of the motion of the observer (the Solar System) on the relativistic thermal and kinematical Sunyaev-Zeldovich effects with the power series approximation in terms of $\theta_{e} \equiv k_{B}T_{e}/mc^{2}$.  We made full use of the Lorentz covariance properties of the problem to obtain solutions with high precision, and confirmed the correctness of the results recently obtained by Chluba, Huetsi, and Sunyaev in the lowest-order of the observer's velocity $\beta_{S}$.  We have given a more general analytic expression for the thermal and kinematical Sunyaev-Zeldovich effects corresponding to the observer's system (the Solar System).  We also found that the effect of the motion of the observer (the Solar System) on the thermal and kinematical Sunyaev-Zeldovich effects is marginally important for present and near-future observations.  However, for the next-generation high-precision observations, we find the effect of the motion of the observer (the Solar System) could be important and should be taken into account for analysis of the observational data.  We also note that the next-generation high-precision kinematic Sunyaev-Zeldovich effect observations will hopefully allow us to determine the cluster velocity perpendicular to the photon propagation vector, as well as the cluster velocity in the direction of the photon propagation vector.

\begin{acknowledgements}

We wish to thank Dr. Chluba for sending their manuscript to us prior to publication of their paper.  We also wish to thank our referee for the helpful suggestions for improving the original manuscript.  This work is financially supported in part by the Grants-in-Aid of the Japanese Ministry of Education, Culture, Sports, Science, and Technology under contracts \#13640245, \#15540293, and \#16540220.

\end{acknowledgements}

\end{document}